\documentclass[harvard]{iss2015}

\usepackage{harvard}
\usepackage{times}
\usepackage{mathptm}
\usepackage{graphicx}
\usepackage{amsmath}
\usepackage{amssymb}
\usepackage{color}

\citationstyle{dcu}

\title{ A $1 + O(1/n)$ APPROXIMATION ALGORITHM FOR TTP(2)}

\authora{Shinji Imahori}
\affiliationa{Faculty of Science and Engineering, Chuo University \\
  1-13-27 Kasuga, Bunkyo, Tokyo 112-8551, Japan \\
  imahori@ise.chuo-u.ac.jp \\}    

\date{\empty}
\def\psbox[#1]#2{\epsfile{file=#2,#1}}

\begin{document}

\maketitle

\begin{abstract}
The traveling tournament problem is a well-known benchmark problem 
of the sports scheduling. 
We propose an approximation algorithm for the traveling tournament problem 
with the constraints such that both the number of consecutive home games 
and that of consecutive away games are at most two (called TTP(2)). 
The approximation ratio of the proposed algorithm is $1 + 24/n$ 
for $n$ teams, which is the first 1 + $O(1/n)$ approximation algorithm  
for TTP(2). 

\end{abstract}

\section{INTRODUCTION}
\label{introduction}

Sports scheduling is an attractive research area in these decades 
\cite{deWerra2006,Trick2011}.  
In the field of sports scheduling, the traveling tournament problem (TTP) 
is a well-known benchmark problem proposed by \citeasnoun{Easton2001}. 
Various approximation algorithms were proposed for the problem 
in the last decade~\cite{Hoshino2012,Imahori2014,Thielen2012,Yamaguchi2011}. 
In the following, some terminology and the problem are introduced. 

Given a set~$T$ of $n$ teams, where $n \geq 4$ and is even, 
a game is specified by an ordered pair of teams. 
Each team has its home venue. 
For any pair of teams $i, j \in T$, $d_{ij} \geq 0$ denotes 
the distance between the home venues of $i$ and~$j$.
Throughout the paper, we assume that 
triangle inequality ($d_{ij} + d_{jk} \geq d_{ik}$) 
and symmetry ($d_{ij} = d_{ji}$) hold. 

A double round-robin tournament is a set of games 
in which every team plays every other team once 
at its home venue (called {\it home} game) and once 
at the home venue of the opponent (called {\it away} game). 
Consequently, $2(n-1)$ slots are necessary to complete 
a double round-robin tournament with $n$ teams. 

Each team stays at its home venue before a tournament 
and then travels to play its games at the chosen venues. 
After a tournament, each team returns to its home venue 
if the team plays an away game at the last slot.  
When a team plays two consecutive away games, 
the team goes directly from the venue of the first opponent 
to that of another opponent without returning to its home venue. 
The traveling distance of a team is defined by the sum of 
distances~$d_{ij}$ if the team travels from the home venue 
of $i$ to the home venue of $j$. 

The objective of TTP is to minimize the total traveling distance, 
which is the sum of traveling distances of $n$ teams. 
Two types of constraints, called {\it no-repeater} and {\it at-most} constraints, 
should be satisfied. 
The no-repeater constraint is that, for any pair of teams $i$ and $j$, 
two games of $i$ and $j$ cannot be held in two consecutive slots.   
The at-most constraint is that, for a given parameter $k$, 
no team plays more than $k$ consecutive home games 
and more than $k$ consecutive away games. 
The present paper considers the case for $k=2$;
the problem is called TTP(2), which is defined as follows. 

\medskip
\noindent
{\bf Traveling Tournament Problem for $k = 2$ (TTP(2))}\\
{\bf Input:\/} A set of teams $T$ and 
		a distance matrix~$D=(d_{ij})$. \\
{\bf Output:\/}
  A double round-robin tournament such that

\noindent
C1. No team plays more than two consecutive away games,

\noindent
C2. No team plays more than two consecutive home games, 

\noindent
C3. The no-repeater constraint is satisfied for all teams, 

\noindent
C4. The total distance traveled by the teams is minimized. 

\medskip 

For this problem, \citeasnoun{Thielen2012} proposed two types 
of approximation algorithms. 
The first algorithm is a 1.5 + $O(1/n)$ approximation algorithm. 
The second one is a 1 + $O(1/n)$ approximation algorithm, 
though it works only for the case with $n = 4m$ teams. 

In this paper, we propose a 1 + $24/n$ approximation algorithm 
for the case with $n = 4m + 2$ teams. 
With the algorithm by \citeasnoun{Thielen2012}, 
we achieve an approximation ratio 1 + $O(1/n)$ for TTP(2).

\section{LOWER BOUNDS}
\label{lower bounds}

In this section, we present the independent lower bound for TTP(2) obtained by 
\citeasnoun{Campbell1976} and another lower bound for analyzing schedules 
generated by our approximation algorithm. 
The basic idea of the independent lower bound is that the optimal trips 
for a team can be obtained by computing a minimum weight perfect matching~$M$
in a complete undirected graph $G$ on the set of teams, 
where the weight of the edge from team $i$ to $j$ given as the distance $d_{ij}$ 
between the home venues of teams $i$ and $j$. 

Let $s(i) := \sum_{j \neq i} d_{ij}$ be the sum of weights of the edges 
between team $i$ and all the other teams $j$, and let $\Delta := \sum_i s(i)$. 
Let $d(M)$ be the weight of a minimum weight perfect matching~$M$ in~$G$. 
Then the traveling distance of team $i$ is at least $s(i) + d(M)$,  
and the total traveling distance is at least 
\begin{equation}
\sum_{i=1}^{n} \left( s(i) + d(M) \right) = \Delta + n \cdot d(M), 
\label{lb1}
\end{equation}
which is called the independent lower bound. 
We note that \citeasnoun{Thielen2012} showed that 
this lower bound cannot be reached in general. 

We introduce another lower bound, which is weaker
than the above independent lower bound but is useful 
to analyze the solutions we generate in the next section. 
Let $d(T)$ be the weight of a minimum spanning tree~$T$ in~$G$. 
Then, for any team $i$, $d(T) \le s(i)$ holds. 
Hence the total traveling distance (i.e., another lower bound) is at least
\begin{equation}
\sum_{i=1}^{n} \left( d(T) + d(M) \right) = n \cdot (d(T) + d(M)).
\label{lb2}
\end{equation} 
It is known that Christofides algorithm \cite{Christofides1976} 
for the traveling salesman problem generates a Hamilton cycle~$C$ in~$G$, 
whose length is at most $d(T) + d(M)$. 


\section{ALGORITHM}
\label{algorithm}

We propose an approximation algorithm for TTP(2) 
for the case with $n = 4m + 2$ teams. 
Our algorithm is similar to the 1 + $O(1/n)$ approximation algorithm 
by \citeasnoun{Thielen2012} for $n = 4m$ teams. 
A key concept of the algorithm is the use of a minimum weight perfect 
matching, a Hamilton cycle computed by Christofides algorithm, 
and the circle method to construct a single round-robin tournament. 

We first compute a minimum weight perfect matching~$M$ 
and a Hamilton cycle~$C$ in the graph $G$. 
By using Christofides algorithm, the length of 
the Hamilton cycle~$C$ is at most $d(M) + d(T)$, 
where $T$ is a minimum weight spanning tree in $G$. 
The $n$ teams are assumed to be numbered such that 
the edges $\langle 1, 2 \rangle, \langle 3, 4 \rangle, \ldots, \langle n-1, n \rangle$ 
form the minimum weight perfect matching~$M$ in~$G$. 
Among possible numberings, we choose a numbering 
with the following three properties. \smallskip \\
(a) \ $s(n-5) + s(n-4) + \cdots + s(n) \le 6\Delta/n$, \\
(b) \ $t(n-7) + t(n-6) \le 12 \Delta / \{n(n-6)\}$, \\
(c) \ teams $2, 4, \ldots, n-8$ are appeared in the Hamilton \\
\hspace*{5mm} cycle~$C$ in this order if the other teams are removed, 
\smallskip \\
where $t(i) := d_{i, n-5} + d_{i, n-4} + \cdots + d_{in}$. 
We note that the existence of a numbering with the above property (a)
comes from an equation 
\[ \Delta = \sum_{i=1}^{n} s(i) = \sum_{i=1}^{n/2} \{s(2i-1) + s(2i)\}, \]
in fact, we choose three pairs $\langle 2i-1, 2i \rangle$ in ascending order  
of the values $\{s(2i-1) + s(2i)\}$ for $\langle n-5, n-4 \rangle$, 
$\langle n-3, n-2 \rangle$ and $\langle n-1, n \rangle$. 
We use the following equation for existence of a numbering with property (b):
\[ \sum_{i=1}^{(n-6)/2} \{t(2i-1) + t(2i)\} = s(n-5) + \cdots + s(n) \le 6\Delta /n. \]

\begin{figure}[tb]
  \centering
  \includegraphics[width=8cm]{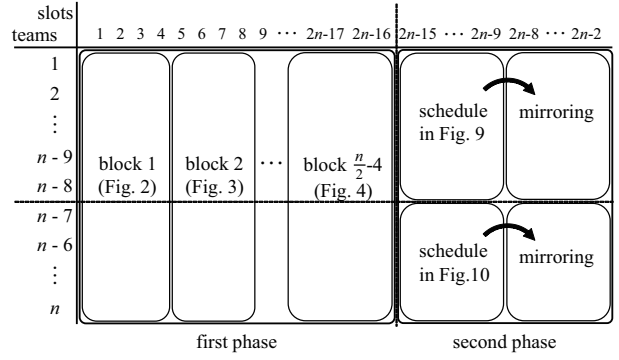}\\
  \caption{Whole image of the schedule we construct}
  \label{whole}
\end{figure}

We then construct a double round-robin tournament. 
As the first phase, we construct a schedule with $2n-16$ slots. 
These $2n-16$ slots are divided into four slots, 
and we call each of them a {\it block}. 
Thus, there are $n/2-4$ blocks in the first phase; 
block 1 contains slots 1, 2, 3, 4, block 2 contains slots 5, 6, 7, 8 and so on.  
As the second phase, we construct schedules of the last 14 slots.
We use the mirroring technique in this phase; a schedule with seven slots 
is constructed and the same schedule is copied by changing the venues. 
Moreover, in this phase, we construct schedules for teams $1, 2, \ldots, n-8$ 
and teams $n-7, n-6, \ldots, n$ independently. 
Fig.~\ref{whole} shows the whole image of the schedule we construct. 

\begin{figure}[htb]
  \centering
  \includegraphics[width=7cm]{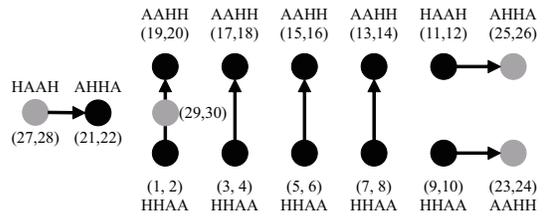}\\
  \caption{Games at the first block for a case with 30 teams}
  \label{first1}
\end{figure}

\begin{figure}[htb]
  \centering
  \includegraphics[width=7cm]{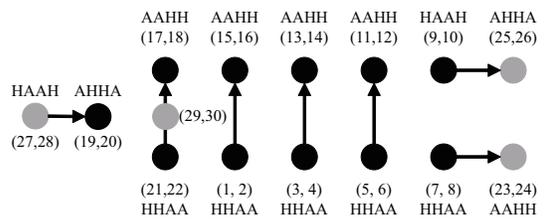}\\
  \caption{Games at the second block for a case with 30 teams}
  \label{first2}
\end{figure}

\begin{figure}[htb]
  \centering
  \includegraphics[width=7cm]{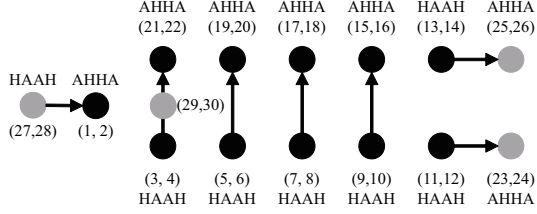}\\
  \caption{Games at the last block for a case with 30 teams}
  \label{first3}
\end{figure}

In the first phase, we apply a scheme inspired by the circle method, 
where a similar idea was also used in \citeasnoun{Thielen2012}. 
As an example, see Fig.~\ref{first1} to Fig.~\ref{first3} for 
a case with $n=30$ teams. 
As displayed in these figures, we put two teams ($2i-1, 2i$) 
for $i = 1, 2, \ldots, n/2$ on one vertex. 

Teams 1 to $n-8$ are put on black vertices in sequence. 
Teams ($n-7, n-6$) are put on the bottom right gray vertex, 
teams ($n-5, n-4$) are put on the top right, 
teams ($n-3, n-2$) are put on the leftmost, 
and teams ($n-1, n$) are put on the gray vertex 
on the leftmost vertical arc.  
Home away patterns of the teams are also displayed in these figures. 
An arc from teams ($i, j$) to ($k, l$) represents 
four games in one block for each team $i, j, k$ and $l$. 
After four games, teams on the black vertices are moved to the next vertices 
as shown in Fig.~\ref{first2}, and then play next four games at the second block. 
This is repeated for $n/2 -4$ times, and Fig.~\ref{first3} shows the last block. 
For all blocks, directions of arcs and teams on the gray vertices are fixed. 
The home away pattern for each vertex is also fixed except for the last block; 
at the last block all the vertices have HAAH or AHHA patterns.  
This modification is done for the compatibility with the second phase. 
We note that there are $n/2 - 4$ possibilities for the initial position 
of teams $(1, 2)$. 
We choose the best one among $n/2 - 4$ possibilities, where the importance 
to choose the best one will be analyzed in the next section. 

\begin{figure}[htb]
  \centering
  \begin{minipage}{0.1\hsize}
  \includegraphics[width=7mm]{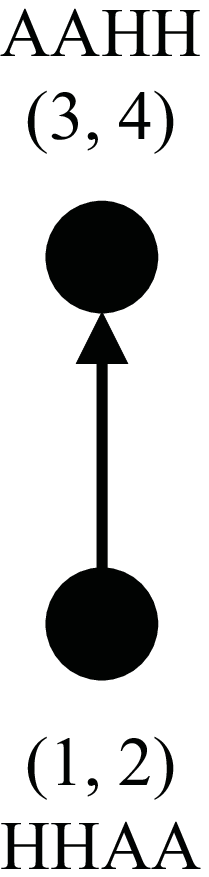}
  \end{minipage}
  \begin{minipage}{0.4\hsize} \ \ 
  \begin{tabular}{r|c@{ \,\,}c@{ \,\,}c@{ \,\,}c@{ \,\,}c}
slots \\
teams \,\,\,& 1 & 2 & 3 & 4\\
\hline
1 & 3H& 4H& 3A& 4A \\
2 & 4H& 3H& 4A& 3A \\
3 & 1A& 2A& 1H& 2H \\
4 & 2A& 1A& 2H& 1H \\
\end{tabular}\\
\end{minipage}
\caption{Games for an arc with HHAA pattern} 
\label{edge1}
\end{figure}
\begin{figure}[htb]
  \centering
  \begin{minipage}{0.1\hsize}
  \includegraphics[width=7mm]{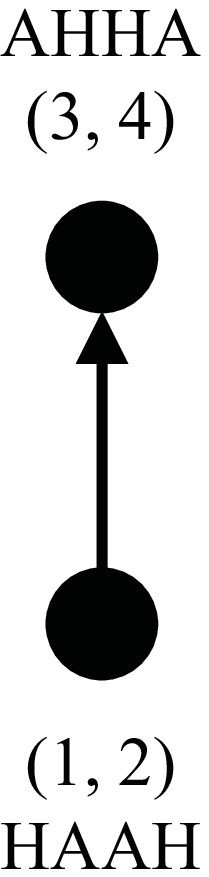}
  \end{minipage}
  \begin{minipage}{0.4\hsize} \ \ 
  \begin{tabular}{r|c@{ \,\,}c@{ \,\,}c@{ \,\,}c@{ \,\,}c}
slots \\
teams \,\,\,& 1 & 2 & 3 & 4\\
\hline
1 & 3H& 4A& 3A& 4H \\
2 & 4H& 3A& 4A& 3H \\
3 & 1A& 2H& 1H& 2A \\
4 & 2A& 1H& 2H& 1A \\
\end{tabular}\\
\end{minipage}
\caption{Games for an arc with HAAH pattern} 
\label{edge2}
\end{figure}
\begin{figure}[htb]
  \centering
  \begin{minipage}{0.15\hsize}
  \includegraphics[width=12mm]{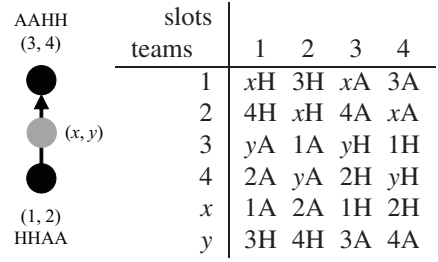}
  \end{minipage}
  \begin{minipage}{0.4\hsize} \ \ 
  \begin{tabular}{r|c@{ \,\,}c@{ \,\,}c@{ \,\,}c@{ \,\,}c}
slots \\
teams \,\,\,& 1 & 2 & 3 & 4\\
\hline
1 & $x$H& 3H& $x$A& 3A \\
2 & 4H& $x$H& 4A& $x$A \\
3 & $y$A& 1A& $y$H& 1H \\
4 & 2A& $y$A& 2H& $y$H \\
$x$ & 1A& 2A& 1H& 2H \\
$y$ & 3H& 4H& 3A& 4A \\
\end{tabular}\\
\end{minipage}
\caption{Games for the leftmost vertical  arc} 
\label{edge3}
\end{figure}
\begin{figure}[htb]
  \centering
  \begin{minipage}{0.15\hsize}
  \includegraphics[width=12mm]{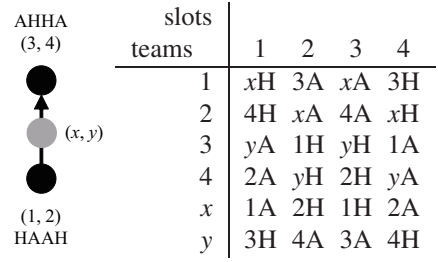}
  \end{minipage}
  \begin{minipage}{0.4\hsize} \ \ 
  \begin{tabular}{r|c@{ \,\,}c@{ \,\,}c@{ \,\,}c@{ \,\,}c}
slots \\
teams \,\,\,& 1 & 2 & 3 & 4\\
\hline
1 & $x$H& 3A& $x$A& 3H \\
2 & 4H& $x$A& 4A& $x$H \\
3 & $y$A& 1H& $y$H& 1A \\
4 & 2A& $y$H& 2H& $y$A \\
$x$ & 1A& 2H& 1H& 2A \\
$y$ & 3H& 4A& 3A& 4H \\
\end{tabular}\\
\end{minipage}
\caption{Games for the leftmost vertical  arc at the last block} 
\label{edge4}
\end{figure}
Here, we describe games for each arc in Fig.~\ref{edge1} to Fig.~\ref{edge4}. 
If one vertex has HHAA and the other has AAHH patterns, 
each team plays games as displayed in Fig.~\ref{edge1}. 
Each number in this table corresponds to the opponent 
and away (home) game is denoted by A (H). 
Fig.~\ref{edge2} shows the games for an arc with HAAH and AHHA patterns, 
this kind of arcs appear in the top right and the leftmost horizontal arcs and 
at the last block. 
We have an exceptional arc in Figs.~\ref{first1} to \ref{first3}, 
the left most vertical arc with an intermediate gray vertex. 
This kind of arc is not necessary for the case with $n=4m$ teams.  
For six teams on this arc, we give four games for each team 
as described in Fig.~\ref{edge3} and Fig.~\ref{edge4}. 

In the second phase, every team plays the remaining games. 
Let $T_1 := \{1, 2, \ldots, n-8 \}$ and 
$T_2 :=  T \setminus T_1$. 
Then, for every pair of teams $i \in T_1$ and $j \in T_2$, 
team~$i$ plays with team~$j$ for two games in the first phase. 
Thus we can construct second phase schedules for $T_1$ and $T_2$ independently. 

\begin{figure}[htb]
\centering
  \includegraphics[width=7cm]{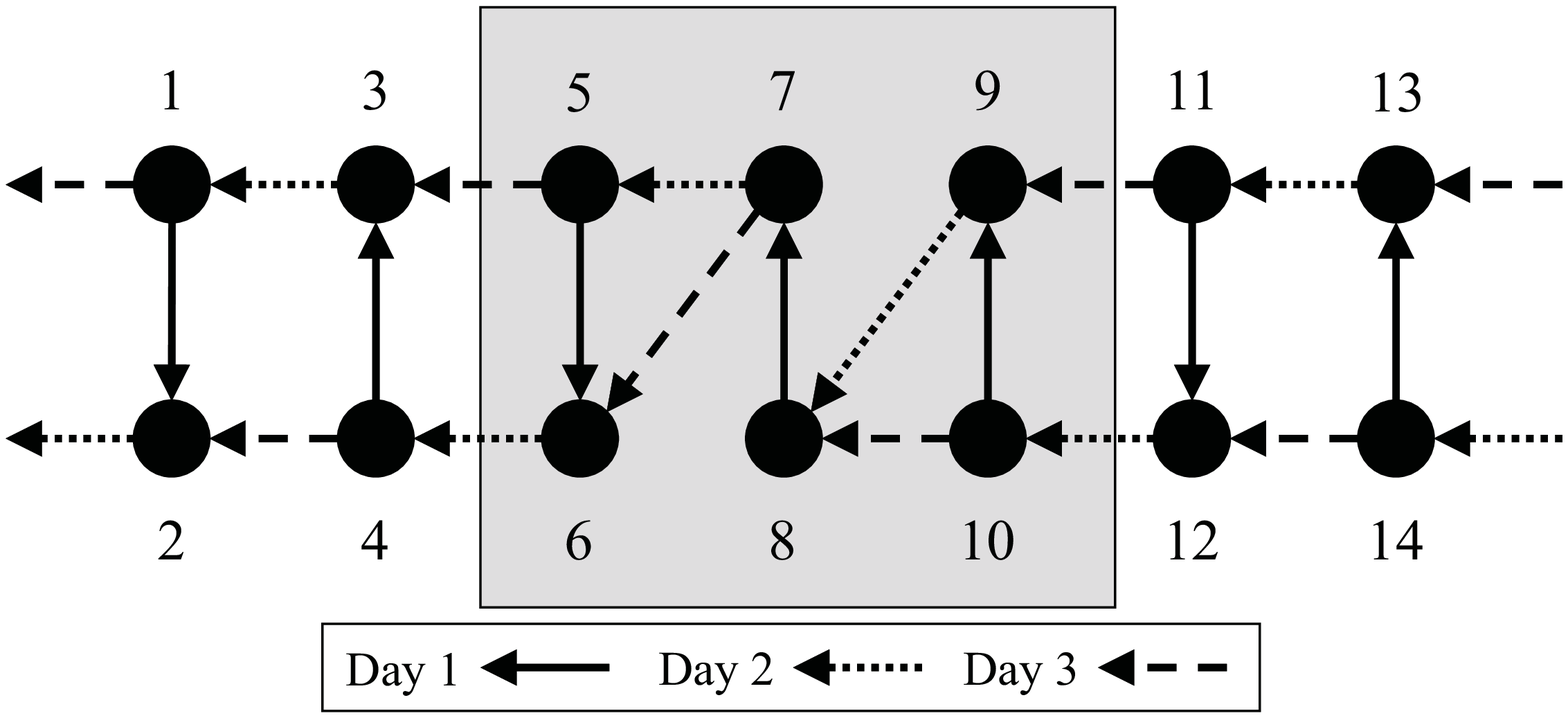} \vspace*{3mm} \\
  \includegraphics[width=7cm]{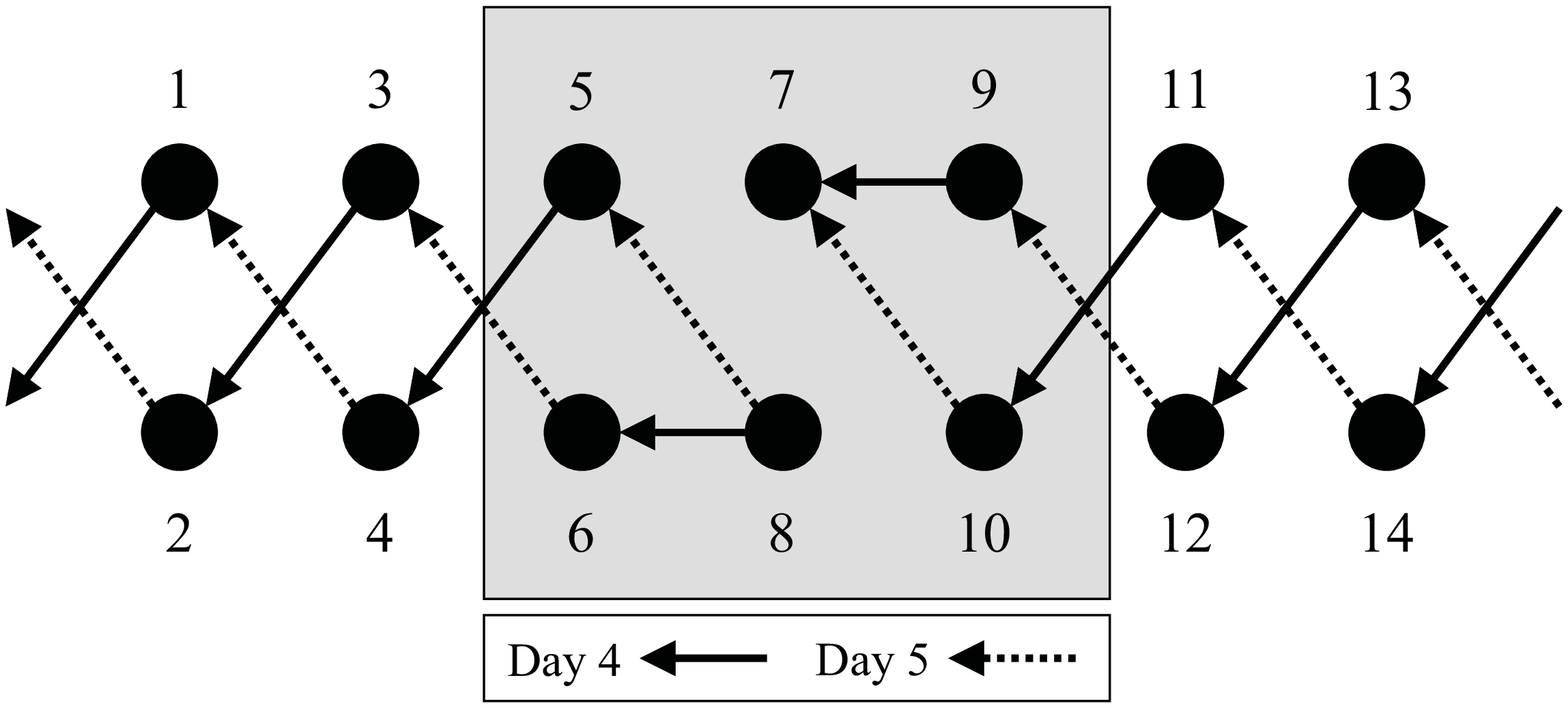} \vspace*{3mm} \\
  \includegraphics[width=7cm]{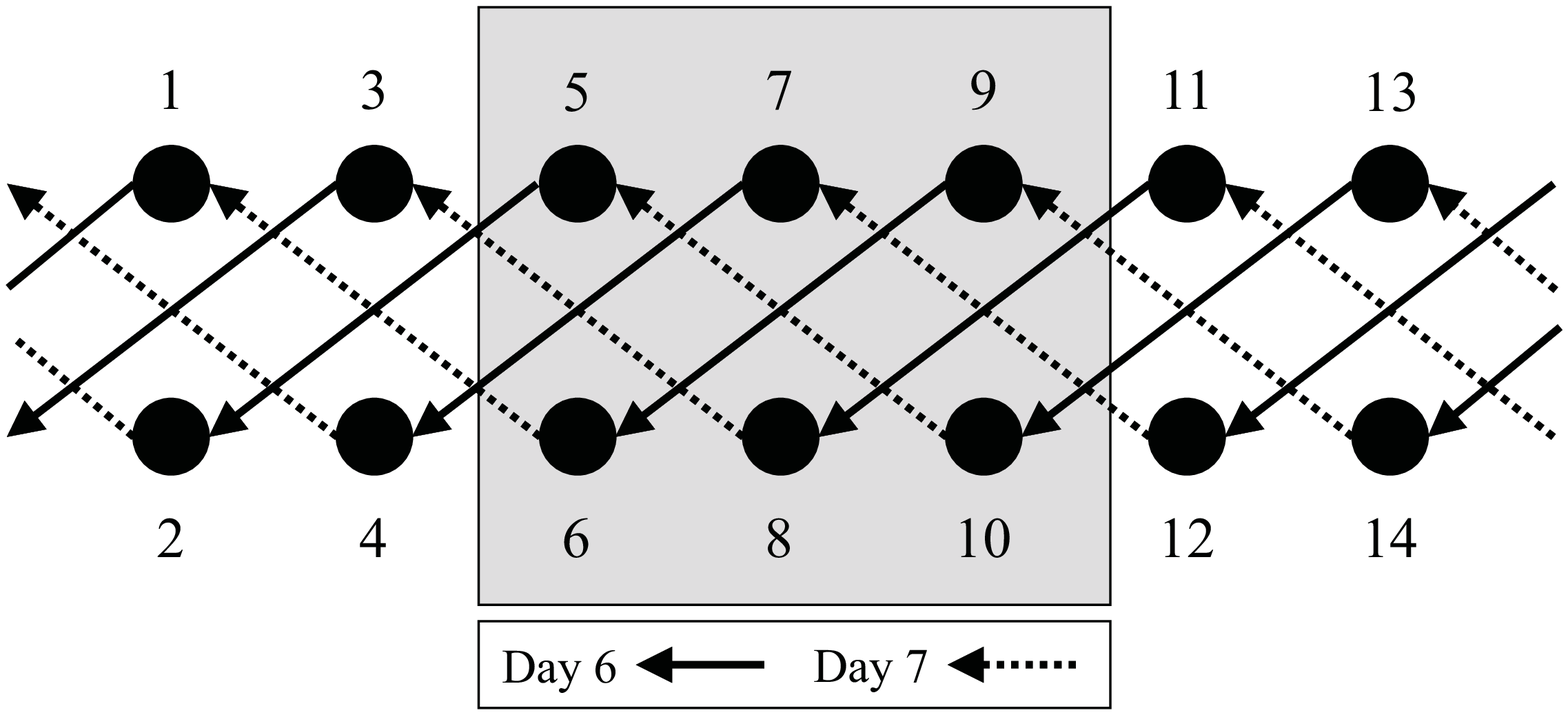}\\
\caption{A schedule for teams $1, 2, \ldots, n-8$ in the second phase
(before the mirroring)} 
\label{second}
\end{figure}
For teams belonging to $T_1$, team $2i-1$ has not played against 
teams $2i-4, 2i-3, 2i-2, 2i, 2i+1, 2i+2, 2i+4$, and 
team $2i$ has not played against 
teams $2i-5, 2i-3, 2i-2, 2i-1, 2i+1, 2i+2, 2i+3$, 
where mod $2n-8$ is applied for team numbers 
(more precisely, 
team $2n-7$ means team 1, 
team 0 means team $2n-8$, 
team $-1$ means team $2n-9$ and so on).  
Fig.~\ref{second} shows how we schedule the remaining games 
for teams in $T_1$. Every arc in this figure corresponds 
to a game at the venue of the team at the head of the arc. 
Apart from the games (of Days 1 to 4) in the gray box, the schedule is 
regularly constructed and can be repeated as many times 
as necessary. 
We note that the number of teams on the upper column 
of Fig.~\ref{second} becomes odd for any number of teams $n=4m+2$, 
and we need a gray box in which the schedule can be irregular. 
The schedule shown in Fig.~\ref{second} is designed  
such as (1) no team plays more than two consecutive home/away 
games in these seven slots, 
(2) every team has HA or AH pattern in Days (slots) 1 and 2, 
and (3) every team has HA or AH pattern in Days 6 and 7. 
By the mirroring technique, we construct the same schedule 
except for the venues of games. 
By concatenating them, the second phase (14 slots) schedule 
for teams belonging to $T_1$ is constructed. 

A schedule for teams in $T_2$ is constructed by 
a single round-robin tournament for eight teams 
with the following properties: 
(1) no team plays more than two consecutive home/away games, 
(2) every team has HA or AH pattern in the first two slots, 
and (3) every team has HH or AA pattern in the first 
and the last slots. 
One example of such single round-robin tournaments 
is displayed in Fig.~\ref{second2}. 
We construct 14 slots schedule for teams in $T_2$ by constructing 
the same (except for the venues) schedule with the mirroring 
technique and concatenating them. 
\begin{figure}[ht]
\centering
\noindent
\begin{tabular}{r|c@{ \,\,}c@{ \,\,}c@{ \,\,}c@{ \,\,}c@{ \,\,}c@{ \,\,}c@{ \,\,}c}
 slots \\
 teams \,\,\,& 1 & 2 & 3 & 4 & 5 & 6 & 7  \\
 \hline
1 & 3H& 4A& 5H& 2A& 6A& 8H& 7H \\
2 & 4H& 3A& 6A& 1H& 5H& 7A& 8H \\
3 & 1A& 2H& 7H& 4A& 8A& 6H& 5A \\
4 & 2A& 1H& 8A& 3H& 7H& 5A& 6A \\
5 & 7H& 8A& 1A& 6H& 2A& 4H& 3H \\
6 & 8H& 7A& 2H& 5A& 1H& 3A& 4H \\
7 & 5A& 6H& 3A& 8H& 4A& 2H& 1A \\
8 & 6A& 5H& 4H& 7A& 3H& 1A& 2A \\
\end{tabular}
\caption{A schedule for teams $n-7, n-6, \ldots, n$ in the second phase
(before the mirroring)} 
\label{second2}
\end{figure}

\section{ANALYSIS OF ALGORITHM}

In this section, we show the feasibility of the schedule we constructed
in Section~3, 
and estimate the approximation ratio of the proposed algorithm. 

\vspace*{-3mm}
\paragraph{Feasibility}
We first show that the constructed schedule in the previous section 
is a double round-robin tournament. 
It is clear that every team plays exactly one game for each slot. 
We check whether every team~$i$ plays with every other team~$j$ 
once at its home and once at $j$'s home venues. 
In the first phase, we use (a modified version of) 
the circle method, which constructs a single round-robin 
tournament (instead of the rightmost part and 
the vertical arc with an intermediate vertex). 
For each arc without an intermediate vertex, 
we assign games as shown in 
Figs.~\ref{edge1} and \ref{edge2}, 
which means that teams $i$ and $j$ have 
two games at $i$'s and $j$'s home venues  
if they are put on the opposite vertices of an arc. 
For the arc with an intermediate vertex 
(see Figs.~\ref{edge3} and \ref{edge4}), some pairs of teams 
play exactly two games as home and away games. 
In the second phase, teams $i$ and $j$ who have not played 
each other play games at $i$'s and $j$'s home venues. 
Thus the schedule is a double round-robin tournament. 

Next we consider the no-repeater constraint. 
If team~$i$ plays with team~$j$ in the first phase, 
two games between $i$ and $j$ appear in an identical block. 
The no-repeater constraint is satisfied for games 
in a block as shown in Figs.~\ref{edge1} to \ref{edge4}. 
If team~$i$ plays with team~$j$ in the second phase, 
two games between $i$ and $j$ do not appear in a consecutive 
two slots since we apply the mirroring technique to single 
round-robin tournaments in Figs.~\ref{second} and \ref{second2}. 

Finally we check whether the schedule satisfies the at-most constraint. 
In each block of the first phase, every team plays two home games 
and two away games. If a team plays home (away) game at the 
last slot of a block, the team plays away (home) game at the 
first slot of the next block. Thus, all teams have at most two 
consecutive home/away games. 
We note that, in the first phase, 
the reason why we put two gray vertices on the rightmost positions 
(two vertices for teams 23 to 26 for the case with $n=30$) 
is that the teams on the rightmost black vertices 
cannot play games each other under both of 
the no-repeater and at-most constraints. 
The reason why we need an intermediate gray vertex 
(for teams 29 and 30) in Fig.~\ref{first1} to Fig.~\ref{first3} 
is that the number of teams $n = 4m + 2$ cannot be divided by four. 
Every team has HA or AH pattern at the end of the first phase. 
This property is also held at the beginning of the second phase.  
Hence, three consecutive home/away games cannot appear 
at the junction of two phases. 
In the second phase, as shown in Fig.~\ref{second} and 
Fig.~\ref{second2}, no team have more than two consecutive 
home/away games. 
Therefore, the constructed double round-robin tournament 
satisfies the at-most constraint. 

\vspace*{-3mm}
\paragraph{Approximation ratio}
We then estimate the total traveling distances of the teams in our 
double round-robin tournament and evaluate the approximation ratio. 
When a team plays two consecutive away games, 
as stated in Section~\ref{introduction}, 
the team is able to go directly from the venue of the first opponent 
to that of another opponent without returning to its home venue 
(with this shortcut, the team can shorten its traveling distance). 
However, in this analysis, we consider that the team returns to 
its home venue unless two consecutive away games are within a block 
in the first phase. 

In the first phase, except for the last block, we consider three types of 
arcs as shown in Figs.~\ref{edge1}, \ref{edge2} and \ref{edge3}. 
Games for the arc in Fig.~\ref{edge1} are the ideal; 
the total traveling distances of the four teams (team 1 to 4) 
in this block is $(d_{13} + d_{14} + d_{23} + d_{24} + d_{31} + d_{32} 
+ d_{41} + d_{42}) + 2 \times (d_{12} + d_{34})$, 
where edges $\langle 1, 2 \rangle$ and $\langle 3, 4 \rangle$ 
are belonging to the minimum weight perfect matching~$M$. 
Traveling distances for games for the arc in Fig.~\ref{edge2} are 
$(d_{13} + d_{14} + d_{23} + d_{24} + d_{31} + d_{32} + d_{41} + d_{42})
+ (d_{31} + d_{32} + d_{41} + d_{42}) + 2 \times (d_{34})$, 
where the second term $(d_{31} + d_{32} + d_{41} + d_{42})$ is surplus. 
This kind of arc appears at the top right and leftmost in Figs.~\ref{first1} 
and \ref{first2}; we assume that teams on the gray vertices 
 (teams 25 to 28 if $n=30$) take this surplus. 
Traveling distances for games for the arc in Fig.~\ref{edge3} are 
$(d_{13} + d_{1x} + d_{24} + d_{2x} + d_{31} + d_{3y} + d_{42} + d_{4y} 
+d_{x1} + d_{x2} + d_{y3} + d_{y4}) + (d_{x3} + d_{x4} + d_{y1} + d_{y2})
+ (d_{12} + d_{34})$. 
Teams on the gray vertex (i.e., teams 29 and 30 for $n=30$) 
take the surplus $(d_{x3} + d_{x4} + d_{y1} + d_{y2})$. 

At the last block of the first phase, we have different home away patterns 
as shown in Fig.~\ref{edge2} and Fig.\ref{edge4}. 
Traveling distances for games for the arc in Fig.~\ref{edge2} are, 
as stated in the previous paragraph, 
$(d_{13} + d_{14} + d_{23} + d_{24} + d_{31} + d_{32} + d_{41} + d_{42}) 
+ (d_{31} + d_{32} + d_{41} + d_{42}) + 2 \times (d_{34})$, 
where the second term $(d_{31} + d_{32} + d_{41} + d_{42})$ is surplus. 
Traveling distances for games for the arc in Fig.~\ref{edge4} are 
$(d_{13} + d_{1x} + d_{24} + d_{2x} + d_{31} + d_{3y} + d_{42} + d_{4y} 
+d_{x1} + d_{x2} + d_{y3} + d_{y4}) + (d_{x3} + d_{x4}) 
+ (d_{1x} + d_{2x} + d_{3y} + d_{4y} + d_{13} + d_{24}) + d_{34}$. 
Team $x$ (team 29 if $n=30$) takes the surplus $d_{x3} + d_{x4}$. 

We now analyze the total traveling distances of the teams in the first phase. 
For each edge belonging to the minimum weight perfect matching~$M$, 
at most two teams use it for one block, and there are $n/2 -4$ blocks. 
Hence the total traveling distances related to the matching~$M$ is 
\begin{equation} 
\left(\frac{n}{2} -4\right) \times 2 \times d(M) 
= (n-8) \cdot d(M).
\label{eq1}
\end{equation}
We then consider the surplus for the last block; 
surpluses for (normal) edges with two vertices and that for the edge 
with an intermediate vertex, 
namely $(d_{31} + d_{32} + d_{41} + d_{42})$ for Fig.~\ref{edge2} 
and $(d_{1x} + d_{2x} + d_{3y} + d_{4y} + d_{13} + d_{24})$ 
for Fig.~\ref{edge4}. 
As stated in Section~\ref{algorithm}, there are $n/2 - 4$ possibilities 
for the initial position of teams on the black vertices. 
If an edge with weight $d_{ij}$ appears as the above surpluses 
for an initial position, the edge cannot appear as surpluses 
for different initial positions. 
Thus the total surpluses for the last slot of the first phase is at most 
$\Delta/2(n/2 - 4)$ on average. Choosing the best initial position, 
the surplus for the last slot is at most 
\begin{equation}
\Delta/2(n/2 - 4) = \Delta/(n - 8).
\label{eq2}
\end{equation} 

In the second phase, we evaluate traveling distances of 
teams 1 to $n-8$ (set $T_1$ with $n-8$ teams) 
and teams $n-7$ to $n$ (set $T_2$) independently. 
We first consider teams in $T_1$ with Fig.~\ref{second}.  
For example, 
team 7 goes to the home venues of teams 4,5,6,8,9,10 and 12. 
Except for single trips from the home venue of team 7 to those of the other teams, 
team 7 takes the surplus $(d_{74} + d_{75} + d_{76} + d_{78} 
+ d_{79} + d_{7,10} + d_{7,12})$. With the triangle inequality, 
it is at most $(d_{56} + 7 \times d_{78} + d_{9,10}) + 
(d_{46} + 3 \times d_{68} + 3 \times d_{8,10} + d_{10,12})$. 
Similarly, team~8 visits the home venues of teams 3,5,6,7,9,10 and 11. 
Except for single trips, 
team 8 takes the surplus $(d_{83} + d_{85} + d_{86} + d_{87} 
+ d_{89} + d_{8,10} + d_{8,11})$, and it is at most 
$(d_{34} + d_{56} + d_{78} + d_{9,10} + d_{11,12}) + 
(d_{46} + 3 \times d_{68} + 3 \times d_{8,10} + d_{10,12})$. 
By considering all the teams in $T_1$, the surplus for teams in $T_1$ 
in the second phase is at most 
\begin{align}
& 14 \times (d_{12} + d_{34} + \cdots + d_{n-9,n-8}) \nonumber \\ 
& + 16 \times (d_{24} + d_{46} + \cdots + d_{n-10,n-8} + d_{n-8,2}) \nonumber \\ 
& \le 14 \times d(M) + 16 \times (d(T) + d(M)). \label{eq3}
\end{align} 

Teams belonging to the set $T_2$ play a double round-robin tournament 
of eight teams in the second phase. 
We evaluate surpluses in this schedule except for the single trips of the teams. 
Among teams in $T_2$, teams $n-5, n-4, \ldots, n$ have already taken 
some surpluses in the first phase. 
We can evaluate the surpluses of those teams $i$ within both of the phases as $s(i)$. 
For the remaining surpluses, team $n-7$ takes $t(n-7)$, team $n-6$ takes $t(n-6)$ 
and we have $2 \times d_{n-7,n-6}$, where edge $\langle n-7, n-6 \rangle$ belongs to $M$.  
The rules (a) and (b) for the numbering of the teams give the following 
inequalities: 
\begin{align} 
& s(n-5) + s(n-4) + \cdots + s(n) \le 6 \Delta / n, \label{eq4} \\
& t(n-7) + t(n-6) \le 12 \Delta / \{n(n-6)\}. \label{eq5}
\end{align}

Finally, we evaluate the total traveling distance of the double round-robin 
tournament we constructed. 
We assume that the number of teams $n$ is at least 30; 
for the case with $n < 30$ teams, it is possible to enumerate all the possible 
schedules and choose an optimal solution in constant time (since 30 is a constant number). 
Under this assumption, we can use the following inequality  
\begin{equation} 
\Delta/(n-8) + 12 \Delta / \{n(n-6)\} \le 2 \Delta/n. 
\label{eq6}
\end{equation}
With the sum of single trips of $n$ teams ($= \Delta$) 
and equations (\ref{eq1}) to (\ref{eq6}), the total traveling distance is at most 
\begin{align} 
& \left(1+\frac{8}{n}\right)\Delta + (n+6) \cdot d(M) + 16 \cdot (d(T) + d(M)) \nonumber \\
\le & \left(1 + \frac{8}{n} \right) (\Delta + n \cdot d(M)) +
\frac{16}{n} (n \cdot  (d(T) + d(M)).  
\end{align} 
By using two kinds of lower bounds (\ref{lb1}) and (\ref{lb2}), 
we have derived a 1 + $24/n$ approximation for TTP(2) 
with $n = 4m + 2 \ (\ge 30)$ teams. 
With the 1 + $16/n$ approximation algorithm for TTP(2) 
with $n = 4m \ (\ge 12)$ teams by \citeasnoun{Thielen2012} 
and a naive enumeration algorithm for the case with a constant 
number of teams  ($n < 30$), we attain the first $1 + O(1/n)$ 
approximation algorithm for TTP(2).

\section{CONCLUSIONS AND FUTURE WORK}

This paper studied an approximation algorithm for  
the traveling tournament problem with constraints that 
the maximum length of home/away consecutive games is two. 
\citeasnoun{Thielen2012} proposed a 1 + $16/n$ 
approximation algorithm for the problem with $n = 4m$ teams. 
In this paper we proposed a 1 + $24/n$ approximation 
algorithm for the case with $n = 4m+2$ teams, this new result 
completes a 1 + $O(1/n)$ approximation algorithm for TTP(2). 

A remained open problem is to reveal the complexity of the problem. 
It was showed by \citeasnoun{Thielen2011}  that 
TTP($k$) is NP-hard for $k \ge 3$. However, the complexity 
of TTP(2) is still open. 
Another direction of future research is a single round-robin tournament 
version of TTP(2). 
It is easy to design a constant factor approximation algorithm 
(because any feasible solution yields a 4 approximation), 
but attaining better results (such as 1 + $O(1/n)$ approximation) 
are future work.




\end{document}